\documentclass[prl,showpacs,twocolumn,preprintnumbers,amsmath,aps]{revtex4}
\usepackage{graphicx,hyperref}
\usepackage{dcolumn}
\usepackage{bm}
\usepackage{subfigure}

\newcommand\gk{\Gamma_k}
\newcommand\rk{R_k(q)}
\newcommand\vp{\varphi}

\newcommand\Min{{\hbox{\begin{tiny}m\end{tiny}}}}

\newcommand\vect[1]{ \boldsymbol{ #1}}
\newcommand\ie{{\it i.e. }}
\newcommand\eg{{\it e.g. }}

\newcommand\etab{\bar \eta}

\begin{document}

\title{Nonperturbative functional renormalization group for random
  field models: the way out of dimensional reduction}

\date{\today}
\author{Gilles Tarjus}
\email{tarjus@lptl.jussieu.fr}
\affiliation{LPTL, Universit\'e Pierre et Marie Curie, bo\^ite 121, 4
  Pl. Jussieu, 75252 Paris c\'edex 05, France} 

\author{Matthieu Tissier}
\email{tissier@lptl.jussieu.fr}
\affiliation{LPTL, Universit\'e Pierre et Marie Curie, bo\^ite 121, 4
  Pl. Jussieu, 75252 Paris c\'edex 05, France} 

\begin{abstract}
  We have developed a non-perturbative functional renormalization
  group approach for the random field $O(N)$ model (RFO(N)M) that
  allows us to investigate the ordering transition in any dimension
  and for any value of $N$ including the Ising case. We show that the
  failure of dimensional reduction and standard perturbation theory is
  due to the non-analytic nature of the zero-temperature fixed point
  controlling the critical behavior, non-analycity which is associated
  with the existence of many metastable states. We find that this
  non-analycity leads to critical exponents differing from the
  dimensional reduction prediction only below a critical dimension
  $d_c(N)<6$, with $d_c(N=1)>3$.
\end{abstract}


\maketitle

After decades of intensive investigation, the nature and properties of
the phase transition in systems with quenched disorder remains a much
debated topic.  Among the controversial issues are the critical
behavior and the phase diagram of the random field Ising model
(RFIM)~\cite{nattermann98}. For the RFIM, the diagnosis is well
established: standard perturbation theory ``exactly'' (\ie to all
orders) predicts an equivalence between the critical behavior of the
RFIM in dimension $d$ and that of the corresponding pure Ising model
in dimension $d-2$ (``dimensional
reduction'')~\cite{parisi79,aharony76}, which contradicts rigorous
results showing that the lower critical dimension of the RFIM is 2 and
not 3~\cite{imbrie84}.  The failure of dimensional reduction (DR) and
conventional perturbation theory is attributed to the existence of a
``complex energy landscape'' characterized by many metastable
states~\cite{parisi84b}. However, limited progress has been made for
developing a theory able to cope with metastability in random field
systems, especially for the physically relevant case of the RFIM in
$d=3$. The most significant step has been the formulation of a
perturbative renormalization group (RG) approach for the RFO(N$>$2)M
near the lower critical dimension $d=4$~\cite{fisher85}. Already at
leading order, the calculation includes an infinite set of marginal
operators near the zero-temperature fixed point controlling the
critical behavior, hence the name functional renormalization group
(FRG).  This perturbative FRG has been further developed to study
equilibrium and out-of-equilibrium properties of elastic manifolds in
disordered medias~\cite{fisher86,chauve00,ledoussal02,balents04}, and
it has been shown that the large scale behavior of these systems is
controlled by non-analytic renormalized actions, with non-analycities
encoding the effect of the many metastable states at zero temperature.

In this letter, we develop a non-perturbative FRG for the RFO(N)M by
combining the ideas of the perturbative FRG for disordered systems
with the formalism of the non-perturbative RG for the effective
average action, based on an exact equation (ERGE)~\cite{berges02}.  We
implement a tractable approximation scheme that allows us to recover
the perturbative results in the relevant limits (most importantly, the
perturbative FRG at first order in $\epsilon=d-4$ for
$N>2$ \cite{fisher85,feldman02}) and to describe, approximately but
non-perturbatively, the ordering transition in the whole $d-N$
diagram, including $N=1$ and $d=3$.  This provides a comprehensive
picture of the critical behavior of random field systems.  We find in
particular that this latter is controlled by a zero-temperature fixed
point at which the renormalized effective average action is
non-analytic (albeit in a more complex way than in the random elastic
manifold case~[\onlinecite{fisher86}--\onlinecite{balents04}]);
however, the DR prediction for the critical exponents breaks down only
below a critical dimension $d_c(N)<6$, with $d_c(N=1)>3$.

We start with the standard effective hamiltonian for the RFO(N)M in
$d$ dimensions with an $N$-component field $\vect\chi(x)$ coupled to a
random field $\vect h(x)$ with zero mean and variance
$\overline{h_\mu(x) h_\nu(y)}=\Delta \delta_{\mu \nu}\delta(x-y)$. For
convenience, we derive the ERGE for disorder averaged functions within
the replica formalism, but it could similarly be obtained by using \eg
the dynamic formulation. We thus consider the ``replicated'' action
\begin{equation}
  \label{eq_action_replicated}
  \begin{split}
    S_n[\{\vect\chi_a\},\{&\vect J_a\}]=\int d^dx \Bigg\{\frac1{2T}
    \sum_{a=1}^n \bigg[ \partial\vect{\chi}_a \cdot
    \partial\vect{\chi}_a\\ + \tau \vect{\chi}_a^2 +&\frac
    u{12}\vect{\chi}_a^4\bigg]- \frac\Delta{2T^2}
    \sum_{a,b=1}^n\vect{\chi}_a.\vect{\chi}_b -\sum_{a=1}^n
    \vect{\chi}_a.\vect{J}_a\Bigg\},
  \end{split}
\end{equation}
where we have introduced sources acting separately on each replica,
which therefore explicitly breaks the permutation symmetry between the
replicas.  One can associate to the above action the thermodynamic
potential $W_n[\{\vect J_a\}]=\log Z_n[\{\vect J_a\}]$ where
$Z_n[\{\vect J_a\}]=\int\prod_{a=1}^n\mathcal
D\vect\chi_a\exp(-S_n[\{\vect \chi_a\},\{\vect J_a\}])$, and its
Legendre transform, the effective action $\Gamma_n
[\{\vect\phi_a\}]=-W_n[\{\vect J_a\}]+\int d^dx\ \sum_{a=1}^n
\vect\phi_a\cdot\vect J_a$ which is the generating functional of the
vertex functions. In the following, we drop the subscript~$n$.

To investigate the phase diagram and critical behavior of the model,
we use an ERGE for the effective average action~\cite{berges02}. An
effective average action $\gk$ at the running scale $k$ is obtained by
integrating out fluctuations with momenta $q\gtrsim k$ via the
introduction of an infra-red (IR) cutoff function $\rk$; $\gk$
continuously interpolates between the bare action,
eq.~(\ref{eq_action_replicated}), at the microscopic scale $k=\Lambda$
and the usual effective action when $k\to 0$. It follows an exact flow
equation,
\begin{equation}
  \label{eq_flot_exact}
  \partial_t\gk[\{\vect\phi_a\}]= 1/2\,\text{Tr}\, \partial_t\rk(
  \boldsymbol{\Gamma}^{\boldsymbol{(2)}}_k[\{\vect\phi_a\},q] 
+ \openone \rk)^{-1}
\end{equation}
where $\partial_t$ is a derivative with respect to $t=\ln(k/\Lambda)$,
$\openone$ is the unit matrix with elements
$(2\pi)^d\delta^{(d)}(q-q')\delta_{ab}\delta_{\mu\nu}$,
$\boldsymbol{\Gamma}^{\boldsymbol{(2)}}_k$ is the tensor formed by the
second functional derivative of $\gk$ with respect to the fields
$\phi_a^\mu(q)$ and $\phi_b^\nu(q')$, and the trace involves an
integration over momenta as well as a sum over replica indices and
$N$-vector coordinates.

Eq.~(\ref{eq_flot_exact}) is a complicated functional
integro-differential equation that cannot be solved exactly but
provides a convenient starting point for non-perturbative
approximation schemes. One such scheme that efficiently deals with the
momentum dependence of the vertex functions is the derivative
expansion~\cite{berges02}.  However, disordered systems require more
because inversion of the matrix involving
$\boldsymbol{\Gamma}^{\boldsymbol{(2)}}_k$ in eq~(\ref{eq_flot_exact})
is a difficult task as far as the replica indices are concerned for
non-integer values of $n$. As in the perturbative FRG
approach~\cite{ledoussal02}, we follow the route that consists in
expanding all functions of the replica fields $\{\vect \phi_a\}$ in
increasing number of free replica sums. Illustrated for the potential
$\mathcal U_k(\{\vect \phi_a\})$ at the running scale $k$ (\ie the
effective average action for uniform fields), this gives:
\begin{equation}
  \label{eq_pot_expansion}
  \begin{split}
    \mathcal U_k(\{\vect\phi_a\})=&\sum_{a=1}^n U_k(\vect\phi_a)-\frac 12
    \sum_{a,b=1}^n V_k(\vect\phi_a,\vect\phi_b)
    +\cdots
  \end{split}
\end{equation}
where $U_k$, $V_k$, $\dots$ are continuous functions of their
arguments and satisfy the replica permutation symmetry.  When all
replica fields are equal, each free replica sum brings a factor of $n$
and the procedure amounts to an expansion in powers of $n$ (with $n\to
0$).

Within this framework, the simplest truncation for $\gk$ that already
contains the main ingredients for a non-perturbative approach of the
RFO(N)M is the following:
\begin{equation}
  \label{eq_trunc_1}
  \begin{split}
    \gk[\{\vect \phi_a\}]=\int& d^dx\Bigg\{\sum_{a=1}^n\Big( \frac 12
    Z_{\Min,k} \partial\vect \phi_a\cdot\partial\vect
    \phi_a\\&+U_k(\vect\phi_a) \Big) -\frac 12
    \sum_{a,b=1}^nV_k(\vect\phi_a,\vect\phi_b)\Bigg\}
  \end{split}
\end{equation}
with one single wave function renormalization for all fields,
$Z_{\Min,k}$, which is defined as the derivative w.r.t.  $q^2$ of
$\gk^{(2)}$ evaluated at $q=0$ for a (non-zero) field configuration
$\vect \phi_\Min$ that minimizes the 1-replica potential $U_k$: this
is the so-called pseudo first-order of the derivative
expansion~\cite{berges02}. With the above truncation that keeps only
the first two terms of the expansion in the free replica sums, the
ERGE for $\gk$, eq.~(\ref{eq_flot_exact}), can be reduced to coupled
partial differential equations for the {\em functions}
$U_k(\vect\phi)$ and $V_k(\vect\phi, \vect\phi')$, whereas a running
anomalous dimension is defined as $\eta_k=-\partial_t\log Z_{\Min,k}$.
The details will be given elsewhere.

To study the critical behavior associated with the ordering transition
and search for fixed points (FP) of the flow equations, we introduce
as usual renormalized dimensionless quantities. However, anticipating
that the putative FP is expected at zero
temperature~\cite{villain84,fisher86b}, it is convenient to make
explicit the flow of a running temperature and the associated
exponent.  For simplicity, let discuss first the RFIM.  We define a
renormalized disorder correlation function $\Delta_k(\phi,
\phi')=\partial_\phi\partial_{\phi'} V_k(\phi, \phi')$ and a
renormalized disorder strength $\Delta_{\Min,k}=\Delta_k(\phi_\Min,
\phi_\Min)$.  A running temperature can now be defined by $T_k=Z_{\Min,k}
k^2 \Delta/(\Lambda^2\Delta_{\Min,k}) $: when $k=\Lambda$, it reduces to
$T_\Lambda= T$ (since from
eq.~(\ref{eq_action_replicated}), $Z_\Lambda=1/T$ and
$\Delta_{\Min,\Lambda}=\Delta/T^2$).  An associated running exponent
is obtained from $\theta_k=\partial_t\log T_k$.  By using the
definition of $\eta_k$, one may alternatively introduce an exponent
$\etab_k=-\theta_k+2+\eta_k$ and compute it from the equation
$\etab_k-2\eta_k=\partial_t\Delta_{\Min,k}$.  Dimensionless quantities
(noted by lower cases) appropriate for looking for a zero temperature
FP are then: $\vp=(T_k Z_k k^{-(d-2)})^{1/2}\phi$,
$u_k(\vp)=T_kk^{-d}U_k(\phi)$, $v_k(\vp,\vp')= T_k^2k^{-d}
V_k(\phi,\phi')$, and $\delta_k(\vp,\vp')= \partial_\vp\partial_{\vp'}
v_k(\vp,\vp')$. The procedure can be extended to the RFO(N)M.  It is
however more convenient in this case to introduce the variables
$\rho=\vect\vp^2/2$ and $z=\vect
\vp\cdot\vect\vp'/(4\rho\rho')^{1/2}$. In scaled form, the flow
equations for $u_k(\rho)$ and $v_k(\rho,\rho',z)$ read (for
simplicity, we drop the subscript $k$ for all quantities but $T_k$):
\begin{equation}
  \label{eq_beta_u}
  \begin{split}    
    \partial_t u= ( 2 - d + &\eta - \etab )u+ \left( d-4 +
      \etab \right) \rho u_\rho \\&+ 2 v_d\big[ (N-1)\delta_T
      l_1^d(w_T)+\delta_L l_1^d(w_L) \big] \\&+ 2 v_d T_k\big[ (N-1)
    l_0^d(w_T)+l_0^d(w_L) \big]
  \end{split}
\end{equation}
\begin{equation}
  \label{eq_beta_v}
  \begin{split}
&\partial_t  v= ( 4 - d + 2\eta  - 2 \etab) v + (   d -4 + \etab) (
  \rho {v_{\rho}} + \rho ' {v_{\rho'}} )  \\&
-  {v_d}\Big\{
{(N-1)}\Big[\left(2\rho{v_{\rho}}- z{v_{z}}  \right) \delta_T
l_2^d({w_T})/\rho+ (2\rho'{v_{\rho'}}\\&- z{v_{z}}  ) \delta_T'
l_2^d({w_T'})/\rho' + v_{z}^2
l_{1,1}^d({w_T},{w_T}')/(2\rho\rho')\Big]+  
\\&{(1-z^2)}
\Big[ {v_{zz}}( \delta_Tl_2^d({w_T})/\rho+  \delta_T'l_2^d({w_T}')/
 \rho')+  2\rho v_{\rho z}^2\\& 
l_{1,1}^d({w_T}',{w_L})/ \rho' + 2\rho 'v_{\rho' z}^2
l_{1,1}^d({w_T},{w_L}')/ \rho- \big(  v_{z}^2\\& 
+2z{v_{z}}{v_{zz}} - {(1-z^2)}v_{zz}^2 \big)l_{1,1}^d({w_T},{w_T}')
/(2\rho\rho') \Big]\\&
+ 2\Big[ \delta_L ( {v_{\rho}} + 2\rho{v_{\rho\rho}} )
l_2^d(w_L) +\delta_L' ( {v_{\rho'}}+ 2\rho ' {v_{\rho'\rho'}})\\&
l_2^d ({w_L}') + 4\rho \rho' v_{\rho\rho'}^2
l_{1,1}^d({w_L},{w_L}')\Big]
\Big\}- T_k{v_d}\\&
\Big\{ {(N-1)} \Big[ ( 2\rho{v_{\rho}}-   z{v_{z}} ) l_1^d({w_T})/\rho
+ ( 2\rho'{v_{\rho'}}- z{v_{z}})\\& 
l_1^d({w_T'})/\rho' \Big] 
+  {(1-z^2)}{v_{zz}}\Big[l_1^d({w_T})/\rho
+ l_1^d({w_T}') /\rho ' \Big]+\\&
  2\Big[\left( {v_{\rho}} +2\rho  {v_{\rho\rho}} \right) l_1^d({w_L})+
( {v_{\rho'}} +2\rho'{v_{\rho'\rho'}} )l_1^d({w_L}') \Big]\Big\}
   \end{split}
\end{equation}
where the indices $\rho,\rho',z$ indicate derivatives w.r.t.
$\rho,\rho',z$, and $v_d^{-1}= 2^{d+1}\pi^{d/2} \Gamma(d/2)$;
$w_{T,L}=w_{T,L}( \rho)$, $w_{T,L}'=w_{T,L}( \rho')$, and similarly
for $\delta_{T,L}$ and $\delta_{T,L}'$, where
$w_T(\rho)=u_{\rho}(\rho)$ and $w_L(\rho)=u_{\rho}(\rho)+ 2\rho\,
u_{\rho\rho}(\rho)$ are the transverse and longitudinal masses,
whereas $\delta_T(\rho)=1/(2\rho) v_{z}(\rho,\rho,1)$ and
$\delta_L(\rho)= 2\rho v_{\rho\rho'}(\rho,\rho,1)$ are the transverse
and longitudinal disorder correlation functions when $\vect \vp=\vect
\vp'$. For brevity we have omitted the arguments $\rho$ and
$\rho,\rho',z$ of all functions. Finally, $l_p^d(w)$ and
$l_{p,p'}^d(w,w')$ are the so-called dimensionless threshold
functions, that essentially encode the non-perturbative effects beyond
the standard one-loop approximation: their definition and properties
are discussed at length in ref.~\cite{berges02}. For the present work,
we choose for IR cutoff function $\rk=Z_{\Min,k}(k^2-q^2)
\Theta(k^2-q^2)$ where $\Theta$ is the Heaviside function.  When $N=1$
and $z=\pm 1$, one recovers the RFIM ($\rho$ and $\rho'$ being used in
place of $\vp$ and $\vp'$).

The above flow equations are supplemented by equations for $\eta_k$
and $\etab_k$. For lack of space, we just give here the equation
obtained for $2\eta_k-\etab_k=-\partial_t\Delta_{\Min,k}$ in the case
of the RFIM:
\begin{equation}
  \label{eq_etab}
  \begin{split}
    &2\eta_k-\etab_k=2v_d\big\{l_4^d(u''_\Min){u'''_\Min}^2-
    4l_3^d(u''_\Min)u'''_\Min \delta'_\Min\\
    &+l_2^d(u''_\Min)(\delta''_\Min+\frac 32 {\delta'_\Min}^2-\frac
    {u'''_\Min}{u''_\Min}\delta_\Min-\frac14\Sigma_\Min)+l_1^d(u''_\Min)
    \frac{{\delta'_\Min}^2}{u''_\Min}\\
    &-T_k[l_2^d(u''_\Min){u'''_\Min}\delta'_\Min- l_1^d(u''_\Min)
    (\frac 12\delta''_\Min-\frac {u'''_\Min}{u''_\Min}\delta'_\Min
    +\frac 12\tilde\Sigma_\Min)]\big\}
  \end{split}
\end{equation}
with $\Sigma(\vp)=\lim_{\vp\to\vp'}(\partial_\vp-\partial_{\vp'})^2
(\delta(\vp,\vp')-\delta(\vp,\vp))^2$ and $\tilde\Sigma(\vp)=
\lim_{\vp\to\vp'}(\partial_\vp-\partial_{\vp'})^2\delta(\vp,\vp')$;
the subscript $m$ indicates that the functions are evaluated for
fields equal to $\vp_\Min$ and primes indicate derivatives w.r.t.
$\vp$. Note the appearance of the ``anomalous'' terms $\Sigma_\Min$
and $T_k \tilde\Sigma_\Min$ that can only differ from zero when a
non-analycity (a ``cusp'') in $\vp-\vp'$ appears in the renormalized
disorder function $\delta(\vp,\vp')$ when $\vp'\to\vp$.  Actually, if
$\delta(\vp,\vp')$ is analytic when $\vp'\to\vp$, the flow equations
for $u(\vp)$ and $\delta(\vp,\vp)$ can be closed by taking from the
beginning the replica symmetric limit: this analytic behavior in the
vicinity of replica symmetry is precisely what is implied by the
standard perturbation theory.  {\em In our formalism, breakdown of DR
  thus implies the emergence of a non-analyticity in the renormalized
  disorder correlation function}.  Note also that if a FP is found to
eqs.  (\ref{eq_beta_u},\ref{eq_beta_v}), and provided that
$\theta=2-\eta+\etab>0$, it is at zero temperature and temperature is
irrelevant (albeit dangerously): indeed, in the vicinity of the FP,
$T_k$ flows to zero as $k^\theta$ when $k\to0$.  In most of the
following, we will consider directly the $T=0$ limit, which allows to
drop all terms proportional to $T_k$ in the above equations.

Because of their structure, the above non-perturbative FRG equations
reproduce all perturbative one-loop results in their range of
validity, in particular the $\epsilon=6-d$ expansion at first order
and the $N=\infty$ limit; a stronger property is that one also
recovers the perturbative FRG equation at first order in
$\epsilon=d-4$ for the RFO(N$>$2)M~\cite{fisher85}; In this case,
$d=4$ being the lower critical dimension, the FP occurs (as for the
pure system) for a value of $\rho_\Min$ that goes to infinity as
$1/\epsilon$.  One can thus organize a systematic expansion in powers
of $1/\rho_\Min$.  The longitudinal mass becomes very large around
$\rho_\Min$, and by using the known asymptotic properties of the
threshold functions for large arguments~\cite{berges02}, one can
derive the flow equations for $\rho_\Min$ (obtained from
eq.~(\ref{eq_beta_u}) and the condition $u'(\rho_\Min)=0$) and for the
function $R(z)=v(\rho_\Min,\rho_\Min,z)/(2\rho_\Min^2)$ (recall that
since $\delta_{T,\Min}=1$ by construction, $R'(1)=1/\rho_\Min$). This
latter reads
\begin{equation}
  \label{eq_flot_r}
  \begin{split}
    \partial_tR&(z)=(\epsilon+2\eta)R(z)-C_4/2\Big\{
    (N-1)\big[{ R'(z)^2}\\
    +&2 R'(1)(2R(z)-zR'(z))\big]+(1-z^2)\big[-
    R'(z)^2\\
    +&(1-z^2) R''(z)^2+2( R'(1)-zR'(z)) R''(z)\big]\Big\}
  \end{split}
\end{equation}
where $C_4=2v_4l_2^4(0)=(16\pi^2)^{-1}$, irrespective of the choice of
the IR cutoff function; in addition, the exponent $\eta$ is given by
$\eta\simeq C_4 R'(1)$. The above equations are identical to the FRG
equations at order $\epsilon$ first derived by
D.~Fisher~\cite{fisher85}.

The main advantage of the non-perturbative FRG that we have developed
is that the mechanism by which DR and conventional perturbation theory
break down can be studied in the whole $d-N$ diagram.  Although we
have not yet obtained the full numerical solutions to
eqs.~(\ref{eq_beta_u},\ref{eq_beta_v}), partial solutions and analyses
lead us to propose the following picture.  (1) Except when $N=\infty$
and $d\geq 6$, the analytic FP found in perturbation theory is never
stable (more precisely once unstable).  (2) The stable FP is
characterized by a renormalized disorder correlation function,
involving 2 fields $\vect \vp$ and $\vect \vp'$, which is non-analytic
near $\vect \vp=\vect \vp'\simeq \vect \vp_\Min$: for instance, in the
RFIM $\delta(\vp,\vp')$ is non-analytic in $(\vp-\vp')$ and for the
RFO(N)M near $d=4$, $R(z)$ is non analytic in $(1-z)$; more generally,
the non-analycity appears in the variable $(\vect \vp-\vect \vp')^2$.
(3) The power exponent characterizing the non-analycity increases
discontinuously with $N$ and $d$: {\em e.g.}, near $d=4$ for $N>2$ there is a
cusp $R'(z)\sim(1-z)^{1/2}$ for $2<N\leq18$, a ``sub-cusp''
$R'(z)\sim(1-z)^{3/2}$ for $18<N\lesssim 18.045$, and so on, and at
large $N$ the exponent varies as $N/2$; a similar trend is observed
with increasing $d$, and in $d=6-\epsilon$, the exponent goes as
$1/\epsilon^2$ (a result quite similar to that of
Feldman~\cite{feldman02}). (4) Only when a cusp is present do the
critical exponents change from their DR value, so that there is a
critical line $d_c(N)$ (or $N_c(d)$) separating the two regions of the
$d-N$ plane in which the exponents are equal ($d>d_c(N)$) or not
($d<d_c(d)$) to the DR predictions.  To locate this line, it is
sufficient to study the flow of the second derivative $\partial^2
v/\partial ((\vect \vp-\vect \vp')^2)^2$ when $\vect \vp=\vect \vp'$
under the assumption of analytic behavior: the appearance of a cusp,
or equivalently of a term in $|\vect \vp-\vect \vp'|^3$ in $v$, is
then signalled by a divergence in the second derivative. To simplify
the analysis, we have used an expansion in powers of the fields
(including all terms up to $\phi^4$) around $\vect \vp_\Min$:
$u=u_2(\rho -\rho_\Min)^2$, $v=2 v_1 (\sqrt{\rho
  \rho'}z-\rho_\Min)+v_2(\rho +\rho '-2\rho_\Min)^2+ v_3(\rho -\rho
')^2+v_4 (\sqrt{\rho \rho'}z-\rho_\Min)^2+ v_5(\sqrt{\rho
  \rho'}z-\rho_\Min)(\rho +\rho '-2\rho_\Min)$, where $v_1=1$ by
construction.  The result is shown in fig.~\ref{fig_d_c_de_n}:
\begin{figure}[tbp]
  \centering
  \includegraphics[width=1. \linewidth]{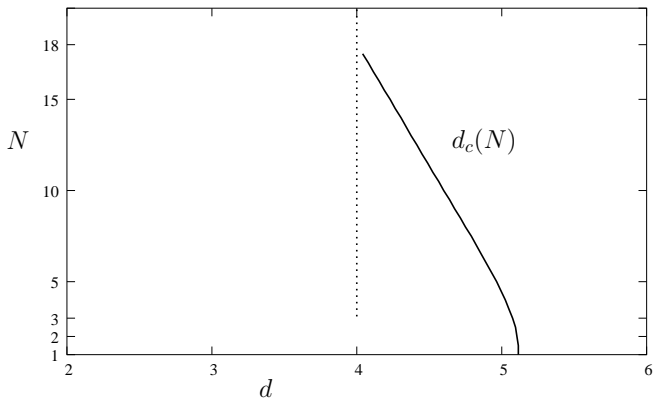}
  \caption{Location of the critical line $d_c(N)$ below which a cusp
  appears and exponents differ from their DR values.}
  \label{fig_d_c_de_n}
\end{figure}
when $d\to4$, one recovers that the critical $N$ is $18$ (see above
and ref.~\cite{fisher85}) and for the
RFIM one finds $d_c(N)\simeq 5.1$ (within the present approximation).
\begin{figure}[b]
  \centering
  \includegraphics[width=.8 \linewidth]{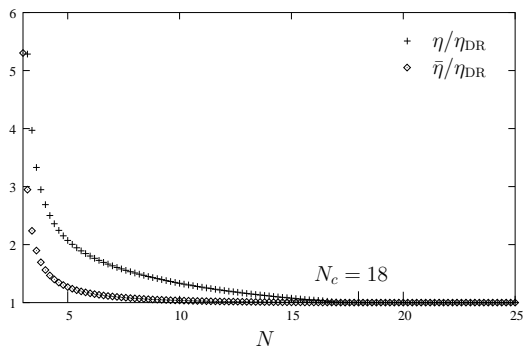}
  \caption{Exponents $\eta$ and $\etab$ (normalized to their
    DR expression) vs $N$ for the
    RFO(N$>$2)M at first order in $\epsilon=d-4$.}
  \label{fig_eta_etab}
\end{figure}
(5) Below $d_c(N)$, the $T=0$ ``cuspy'' FP is associated with critical
exponents differing from their DR value; the simplest illustration is
for the RFO(N$>$2)M at first order in $\epsilon=d-4$: see
fig.~\ref{fig_eta_etab} where we display $\eta$ and $\etab$ normalized
by their DR value $\etab=\eta=\epsilon/(N-2)$ as a function of $N$
(for $N=3,4,5$, they agree with those of ref.~\cite{feldman02}). (6)
For any finite temperature $T_k$, the cusp is rounded: this can be
inferred, \eg, from the term $T_k \tilde\Sigma_\Min$ in
eq.~(\ref{eq_etab}) that must stay finite as one approaches the FP. As
in the random elastic manifold problem~\cite{chauve00,balents04},
temperature prevents the flow from being non-analytic at any finite
scale and one may expect that the rounding of the cusp involves a
boundary layer as $T_k\to 0$ and one approaches the FP.

The present non-perturbative FRG approach of random field systems
provides a consistent and global picture of the critical behavior
associated with the ferromagnetic ordering transition. The failure of
DR and standard perturbation theory comes from the existence of many
metastable states, but the mechanism by which this occurs is rather
subtle: metastability results from an interplay between ferromagnetic
ordering and disorder, and it plays a role at large scale because the
fixed point occurs at $T=0$.  This interplay leads to an effective
renormalized {\em random potential} (beyond the bare random field
term) that displays many minima.  The renormalized disorder
correlation function, which is the second derivative of the second
cumulant of this random potential, acquires a non-analycity (a cusp in
low enough $d$) as it flows to the $T=0$ fixed point.  The physics of
such cusps has been discussed in the context of random elastic
manifolds~\cite{ledoussal02,fisher86,balents96,chauve00};
but in the present case the cusp only occurs when the system is in the
vicinity of the minimum of the {\em non-random potential}. Work is in
progress to obtain the full numerical solution of the flow equations
and compute the critical exponents, as well as investigate the
connection of the present theory with replica symmetry breaking
approaches~\cite{mezard92,brezin01} and possible formation of bound
states~\cite{brezin01,parisi02}.

LPTL is UMR 7600 at the CNRS.


\begin{thebibliography}
\expandafter\ifx\csname natexlab\endcsname\relax\def\natexlab#1{#1}\fi
\expandafter\ifx\csname bibnamefont\endcsname\relax
  \def\bibnamefont#1{#1}\fi
\expandafter\ifx\csname bibfnamefont\endcsname\relax
  \def\bibfnamefont#1{#1}\fi
\expandafter\ifx\csname citenamefont\endcsname\relax
  \def\citenamefont#1{#1}\fi
\expandafter\ifx\csname url\endcsname\relax
  \def\url#1{\texttt{#1}}\fi
\expandafter\ifx\csname urlprefix\endcsname\relax\def\urlprefix{URL }\fi
\providecommand{\bibinfo}[2]{#2}
\providecommand{\eprint}[2][]{\url{#2}}

\bibitem[{\citenamefont{Nattermann}(1998)}]{nattermann98}
\bibinfo{author}{\bibfnamefont{T.}~\bibnamefont{Nattermann}}, in
  \emph{\bibinfo{booktitle}{Spin glasses and random fields}}
  edited by \bibinfo{editor}{\bibfnamefont{A.~P.} \bibnamefont{Young}}
  (\bibinfo{publisher}{World scientific, Singapore},
  \bibinfo{year}{1998}), p. \bibinfo{pages}{277}.

\bibitem{aharony76}
{A.}~{Aharony} {\it et~al}., Phys. Rev. Lett., \textbf{{37}},
  {1364} ({1976}). G.~Grinstein, Phys. Rev. Lett., \textbf{{37}},
  {944} ({1976}).

\bibitem[{\citenamefont{Parisi and Sourlas}(1979)}]{parisi79}
\bibinfo{author}{\bibfnamefont{G.}~\bibnamefont{Parisi}} \bibnamefont{and}
  \bibinfo{author}{\bibfnamefont{N.}~\bibnamefont{Sourlas}},
  \bibinfo{journal}{Phys. Rev. Lett.} \textbf{\bibinfo{volume}{43}},
  \bibinfo{pages}{744} (\bibinfo{year}{1979}).

\bibitem{imbrie84}
J.~Z.~Imbrie, Phys. Rev. Lett. \textbf{{53}}, {1747} ({1984}).
J.~Bricmont and A.~Kupianen, {Phys. Rev. Lett.} \textbf{{59}},
 {1829} ({1987}).

\bibitem[{\citenamefont{Parisi}(1984)}]{parisi84b}
\bibinfo{author}{\bibfnamefont{G.}~\bibnamefont{Parisi}}, in
  \emph{\bibinfo{booktitle}{Proceedings of Les Houches 1982, Session XXXIX}},
  edited by \bibinfo{editor}{\bibfnamefont{J.~B.} \bibnamefont{Zuber}}
  \bibnamefont{and} \bibinfo{editor}{\bibfnamefont{R.}~\bibnamefont{Stora}}
  (\bibinfo{publisher}{North Holland, Amsterdam}, \bibinfo{year}{1984}), p.
  \bibinfo{pages}{473}.

\bibitem[{\citenamefont{Fisher}(1985)}]{fisher85}
\bibinfo{author}{\bibfnamefont{D.~S.} \bibnamefont{Fisher}},
  \bibinfo{journal}{Phys. Rev. B} \textbf{\bibinfo{volume}{31}},
  \bibinfo{pages}{7233} (\bibinfo{year}{1985}).

\bibitem[{\citenamefont{Feldman}(2002)}]{feldman02}
\bibinfo{author}{\bibfnamefont{D.~E.} \bibnamefont{Feldman}},
  \bibinfo{journal}{Phys. Rev. Lett.} \textbf{\bibinfo{volume}{88}},
  \bibinfo{pages}{177202} (\bibinfo{year}{2002}).

\bibitem[{\citenamefont{Fisher}(1986{\natexlab{a}})}]{fisher86}
\bibinfo{author}{\bibfnamefont{D.~S.} \bibnamefont{Fisher}},
  \bibinfo{journal}{Phys. Rev. Lett.} \textbf{\bibinfo{volume}{56}},
  \bibinfo{pages}{1964} (\bibinfo{year}{1986}{\natexlab{a}}).
  
\bibitem[{\citenamefont{Chauve et~al.}(2000)\citenamefont{Chauve,
      Giamarchi, and {Le Doussal}}}]{chauve00}
  \bibinfo{author}{\bibfnamefont{P.}~\bibnamefont{Chauve}} {\it et
    al.}, \bibinfo{journal}{Phys. Rev. B}
  \textbf{\bibinfo{volume}{62}}, \bibinfo{pages}{6241}
  (\bibinfo{year}{2000}).

\bibitem[{\citenamefont{{Le Doussal} and Wiese}(2002)}]{ledoussal02}
\bibinfo{author}{\bibfnamefont{P.}~\bibnamefont{{Le Doussal}}}
  \bibnamefont{and} \bibinfo{author}{\bibfnamefont{K.~J.} \bibnamefont{Wiese}},
  \bibinfo{journal}{Phys. Rev. Lett.} \textbf{\bibinfo{volume}{89}},
  \bibinfo{pages}{125702} (\bibinfo{year}{2002});
  \bibinfo{journal}{Phys. Rev. B} \textbf{\bibinfo{volume}{68}},
  \bibinfo{pages}{174202} (\bibinfo{year}{2003}).

\bibitem[{\citenamefont{Balents and {Le Doussal}}(2004)}]{balents04}
\bibinfo{author}{\bibfnamefont{L.}~\bibnamefont{Balents}} \bibnamefont{and}
  \bibinfo{author}{\bibfnamefont{P.}~\bibnamefont{{Le Doussal}}},
  \bibinfo{journal}{Europhys. Lett.} \textbf{\bibinfo{volume}{65}},
  \bibinfo{pages}{685} (\bibinfo{year}{2004}).

\bibitem[{\citenamefont{Berges et~al.}(2002)\citenamefont{Berges, Tetradis, and
  Wetterich}}]{berges02}
\bibinfo{author}{\bibfnamefont{J.}~\bibnamefont{Berges}} {\it et~al.},
  \bibinfo{journal}{Phys. Rep.} \textbf{\bibinfo{volume}{363}},
  \bibinfo{pages}{223} (\bibinfo{year}{2002}).

\bibitem[{\citenamefont{Villain}(1984)}]{villain84}
\bibinfo{author}{\bibfnamefont{J.}~\bibnamefont{Villain}},
  \bibinfo{journal}{Phys. Rev. Lett.} \textbf{\bibinfo{volume}{52}},
  \bibinfo{pages}{1543} (\bibinfo{year}{1984}).

\bibitem[{\citenamefont{Fisher}(1986{\natexlab{b}})}]{fisher86b}
\bibinfo{author}{\bibfnamefont{D.~S.} \bibnamefont{Fisher}},
  \bibinfo{journal}{Phys. Rev. Lett.} \textbf{\bibinfo{volume}{56}},
  \bibinfo{pages}{416} (\bibinfo{year}{1986}{\natexlab{b}}).
\bibinfo{author}{\bibfnamefont{A.~A.} \bibnamefont{Middleton}}
  \bibnamefont{and} \bibinfo{author}{\bibfnamefont{D.~S.}
  \bibnamefont{Fisher}}, \bibinfo{journal}{Phys. Rev. B}
  \textbf{\bibinfo{volume}{65}}, \bibinfo{pages}{134411}
  (\bibinfo{year}{2002}).

\bibitem[{\citenamefont{Balents et~al.}(1996)\citenamefont{Balents, Bouchaud,
  and M\'ezard}}]{balents96}
\bibinfo{author}{\bibfnamefont{L.}~\bibnamefont{Balents}} {\it et~al.},
  \bibinfo{journal}{J. Phys. I (Paris)} \textbf{\bibinfo{volume}{6}},
  \bibinfo{pages}{1007} (\bibinfo{year}{1996}).

\bibitem[{\citenamefont{Mezard and Young}(1992)}]{mezard92}
\bibinfo{author}{\bibfnamefont{M.}~\bibnamefont{Mezard}} \bibnamefont{and}
  \bibinfo{author}{\bibfnamefont{A.~P.} \bibnamefont{Young}},
  \bibinfo{journal}{Europhys. Lett.} \textbf{\bibinfo{volume}{18}},
  \bibinfo{pages}{653} (\bibinfo{year}{1992}).

\bibitem[{\citenamefont{Br\'ezin and {De Dominicis}}(2001)}]{brezin01}
\bibinfo{author}{\bibfnamefont{E.}~\bibnamefont{Br\'ezin}} \bibnamefont{and}
  \bibinfo{author}{\bibfnamefont{C.}~\bibnamefont{{De Dominicis}}},
  \bibinfo{journal}{Eur. Phys. J. B} \textbf{\bibinfo{volume}{19}},
  \bibinfo{pages}{467} (\bibinfo{year}{2001}).

\bibitem[{\citenamefont{Parisi and Sourlas}(2002)}]{parisi02}
\bibinfo{author}{\bibfnamefont{G.}~\bibnamefont{Parisi}} \bibnamefont{and}
  \bibinfo{author}{\bibfnamefont{N.}~\bibnamefont{Sourlas}},
  \bibinfo{journal}{Phys. Rev. Lett.} \textbf{\bibinfo{volume}{89}},
  \bibinfo{pages}{257204} (\bibinfo{year}{2002}).

\end{thebibliography}
\end{document}